# Discovery of microscopic electronic inhomogeneity in the high-$T_c$ superconductor $Bi_2Sr_2CaCu_2O_{8+x}$

S.H. Pan[*], J.P. O'Neal[*], R.L. Badzey[*], C. Chamon[*], H. Ding[†], J.R. Engelbrecht[†], Z. Wang[†], H. Eisaki[‡§], S. Uchida[‡], A.K. Gupta[∥], K.-W. Ng[∥], E.W. Hudson[¶#], K.M. Lang[¶], and J.C. Davis[¶]

[*] *Department of Physics, Boston University, Boston, MA 02215, USA*
[†] *Department of Physics, Boston College, Chestnut Hill, MA 02467, USA*
[‡] *Department of Superconductivity, University of Tokyo, Tokyo, Japan*
[∥] *Department of Physics and Astronomy, University of Kentucky, Lexington, KY 40506-0055, USA*
[¶] *Department of Physics, University of California, Berkeley, CA 94720, USA*
[§] *Present address: Department of Applied Physics, Stanford University, Stanford, CA 94305, USA.*
[#] *Present address: National Institute of Standards and Technology, Gaithersburg, MD 20899, USA.*

**Correspondence and requests for materials should be addressed to S.H. Pan, (e-mail: shpan@bu.edu)**

**The parent compounds of the copper oxide high-$T_c$ superconductors are unusual insulators. Superconductivity arises when they are properly doped away from stoichiometry[1]. In $Bi_2Sr_2CaCu_2O_{8+x}$, superconductivity results from doping with excess oxygen atoms, which introduce positive charge carriers (holes) into the $CuO_2$ planes, where superconductivity is believed to originate. The role of these oxygen dopants is not well understood, other than the fact that they provide charge carriers. However, it is not even clear how these charges distribute in the $CuO_2$ planes. Accordingly, many models of high-$T_c$ superconductors simply assume that the charge carriers introduced by doping distribute uniformly, leading to an electronically homogeneous system, as in ordinary metals. Here we report the observation of an electronic inhomogeneity in the high-$T_c$ superconductor $Bi_2Sr_2CaCu_2O_{8+x}$ using scanning tunnelling microscopy/spectroscopy. This inhomogeneity is manifested as spatial variations in both the local density of states spectrum and the superconducting energy gap. These variations are correlated spatially and vary on a surprisingly short length scale of ~ 14 Å. Analysis suggests that the inhomogeneity observed is a consequence of proximity to a Mott insulator resulting in poor screening of the charge potentials associated with the oxygen ions left behind in the BiO plane after doping. Hence this experiment is a direct probe of the local nature of the superconducting state, which is not easily accessible by macroscopic measurements.**

We have carried out extensive low-temperature scanning tunnelling microscopy/spectroscopy (STM/S) studies on optimally doped $Bi_2Sr_2CaCu_2O_{8+x}$. Technical details of the experiments have been reported previously[2,3]. Various single crystal samples fabricated with different techniques (directional-flux-solidification or floating-zone) have been used to ensure the generality of the observations. Some crystals are nominally pure and some are deliberately doped with a very dilute concentration of impurity atoms (Zn or Ni). Remarkably, we have consistently observed a particular type of electronic inhomogeneity regardless of the differences between the individual crystals. This inhomogeneity is manifested as spatial variations in the local density of states (LDOS) spectrum, in the low-energy spectral weight, and in the magnitude of the superconducting energy gap.

As an example, Fig. 1a presents a topographic image obtained on a pure $Bi_2Sr_2CaCu_2O_{8+x}$ crystal grown by the flux-solidification method. The image reveals the atomically-resolved crystal structure of the BiO plane accompanied by the well-known incommensurate structural modulations. In addition, an inhomogeneous background is also discernable. After removing the contrast associated with the above-mentioned topological structures by Fourier filtering, this inhomogeneous background is clearly visible. Assuming that the tunnelling matrix element has no spatial dependence, the resulting image shown in Fig. 1b can be approximated as a map displaying the variation of the integrated LDOS. With spatially resolved spectroscopy, we find that the local tunnelling spectrum also varies from location to location. Remarkably, we discover that the magnitude of the superconducting gap extracted from the local tunnelling spectrum varies spatially as well, instead of exhibiting the single-valued nominal gap of ~ 40 meV usually observed by tunnelling in crystals with optimal oxygen doping[4,5]. Surprisingly, both the integrated LDOS and the superconducting gap magnitude vary on an apparently much shorter length scale than those in the phenomena observed in earlier experiments[6,7]. These are dramatic observations compared to the case of conventional BCS superconductors, such as Nb, where the integrated LDOS, the tunnelling spectrum, and the superconducting gap are spatially homogeneous.

What is the origin of such an inhomogeneity? Is it intrinsic or caused by impurities such as crystal defects or excess elements that limit the sample quality? Figure 2 is an example of the STM/S results obtained on high quality, floating-zone-grown single crystals deliberately doped with a very dilute concentration (0.2%) of Zn atoms. The inhomogeneous background observed in these crystals (Fig 2a) is of the same type as that seen in Fig. 1b. This suggests that the inhomogeneity does not originate from impurities. To support this point, we identified the locations of the Zn impurities from a zero bias conductance map, which is taken simultaneously with the LDOS map at the same location[3]. Using cross-correlation analysis, we found no correlation between the intensity of the integrated LDOS and the locations of the Zn impurities. Furthermore, the length scale of the spatial variation observed in the integrated LDOS map is much smaller than the average impurity spacing. Therefore, we conclude that the inhomogeneity observed in the integrated LDOS is not induced by impurities, but rather is intrinsic in nature.

Spatial variations of the tunnelling spectrum and of the superconducting gap, similar to those observed in the pure sample, are observed in this impurity-doped sample (Fig. 2b). Statistical analysis of Fig. 2a shows that the integrated LDOS has a Gaussian distribution, displayed in Fig. 2c. Similarly, analysis of Fig. 2b shows that the gap ranges from 25 meV to 65 meV and exhibits a Gaussian distribution (42 meV mean; ~20 meV FWHM) as shown in Fig. 2d. It is interesting that the *average* gap value is very similar to the single value reported on optimally doped $Bi_2Sr_2CaCu_2O_{8+x}$ in earlier tunnelling measurements[4,5].

In addition to the similarity of their statistical distributions, a strong spatial correlation between the integrated LDOS and the superconducting gap can be seen readily by recognizing similar patterns in the two maps (Fig. 2a and Fig. 2b), which are obtained at the same location simultaneously. As illustrated in Fig. 2e-g, auto-correlation analysis on both maps shows similar decay lengths ξ of approximately 14 Å and cross-correlation shows a pronounced Gaussian peak, confirming the local nature of the variations and the strong correlation of these two quantities. Furthermore, the line-shape of the local tunnelling spectrum also correlates with the integrated LDOS. As the gap and the integrated LDOS variations are correlated and the energy gap variation cannot be attributed to a tunnelling matrix effect, the inhomogeneity in the integrated LDOS we observe is most likely intrinsic to the electronic structure and not due to a spatially varying matrix element.

Spatial variation of the energy gap and its correlation with the LDOS can be seen in greater detail in Fig. 3. Clearly, the spectra obtained at points with larger integrated LDOS values exhibit higher differential conductance, smaller gap values, and sharper coherence peaks. More significantly, these STM/S results resemble planar junction spectra obtained on samples with different oxygen doping concentrations[4,5] — the STM/S spectra obtained at positions with higher values of integrated LDOS resemble the planar junction spectra obtained on samples with higher oxygen doping concentrations. These observations naturally lead one to relate the magnitude of the integrated *local* DOS to the *local* oxygen doping concentration. On a macroscopic scale, such a correspondence is expected for a doped Mott insulator where doping adds spectral weight near the Fermi energy and provides the carriers necessary to transform the insulating compound into a superconductor, as observed previously by photoemission[8]. The idea of *local doping concentration* (LDC) extends this picture to microscopic scales. In this scenario, the charge potentials of oxygen dopants alter the local electronic environment provided that the screening of these potentials is weak on the scale of the inter-carrier distance. For doped Mott insulators, this condition is satisfied by a relatively large inter-carrier distance and strong electronic correlation. A detailed theoretical analysis of this scenario will be discussed elsewhere.

In Figure 4, we present a scatter-plot of the magnitude of the energy gap versus the value of the integrated LDOS. For comparison, the doping dependence of the gap value obtained by angle-resolved photoelectron spectroscopy (ARPES) is superimposed on the plot. Note that STM is a real space, local probe, while ARPES resolves in momentum space but averages in real space over a macroscopic spot. The similarity in the linear behaviour of these two complimentary measurements is remarkable, and provides further support for the concept of LDC in this strongly correlated system.

The existence of such microscopic inhomogeneities should have many important consequences on the quasiparticle properties that are accessible by macroscopic measurements. For example, unexpectedly broad peaks in an earlier neutron scattering measurement on $Bi_2Sr_2CaCu_2O_{8+x}$ can possibly be explained by the presence of inhomogeneity in the bulk[9]. However, it is more important to compare our results to a complementary technique such as ARPES, which measures the same electronic excitations in momentum space under similar experimental conditions. In doing so, we find that taking a spatial average of STM results yields the same maximum gap value and the same doping dependence of the gap as obtained by ARPES. In addition, the large intrinsic width (~ 20 meV) of the coherence peak measured by ARPES[10,11] near the antinodes is consistent with the 20 meV FWHM distribution of the superconducting gap shown in Fig. 2d. In light of the electronic disorder observed by this STM experiment, the discrepancy between the nodal and anti-nodal quasiparticle mean-free-paths observed by ARPES can now be reconciled. Near the antinodes, gap disorder can significantly scatter the quasiparticles at the gap edge. Indeed, the peak in the momentum distribution curve (MDC) near the antinodes has a large width of ~0.08 Å$^{-1}$ [12], corresponding to a short mean-free-path of ~ 25 Å. In the nodal direction, however, ARPES measures a mean-free-path of ~100 Å [13,14]. The quasiparticles of the d-wave superconductor in this direction are less affected by the gap disorder since the latter amounts to velocity disorder in the dispersion, which is not effective at scattering Dirac particles. Thus, the mean-free-path of the nodal quasiparticles will be limited primarily by elastic scattering from the potential disorder. Given that the oxygen dopants responsible for this potential disorder do not reside in the $CuO_2$ plane, scattering is relatively weak. This can result in a longer mean-free-path and the measured 100 Å length scale is therefore quite reasonable. As scattering into all angles is involved in ARPES measurements, while transport is dominated by large-angle scattering of the nodal quasiparticles, our results may also reconcile the discrepancy between the mean-free-path measured by transport[15] as compared to that measured by ARPES. Taking 14 Å as the length scale over which the disorder potential varies significantly, we deduce that the dominant elastic scattering process is limited by the wavevector q = 1/14 Å$^{-1}$. The scattering angle given by $sin(\theta/2)=q/2k_F$ is indeed quite small ~ 5°, which provides a possible explanation as to why the transport mean-free-path can be much longer than that measured by ARPES.

Discussion of our observations can also be extended to more fundamental issues, such as the coherence of the superconducting state. The coexistence of a high superconducting transition temperature with such a microscopic inhomogeneity implies that the superconducting coherence length is shorter than the mean-free-path. Our measured gap correlation length, $\xi \sim 14$ Å, sets the length scale for the superconducting pair size in optimally doped $Bi_2Sr_2CaCu_2O_{8+x}$. It is interesting that by evaluating the BCS expression $\xi_0 = \eta v_F/\pi\Delta$, taking $\eta v_F = 1.6$ eVÅ from band dispersion near the nodes[14,16] and the averaged gap at optimal doping as $\Delta = 0.04$ eV, we obtain $\xi_0 \sim 13$ Å, in good agreement with the correlation decay length $\xi$ obtained from our experiment. Yet it is more intriguing that $\xi$ appears shorter than the experimental in-plane superconducting coherence length $\xi_{ab} \sim 22$-$27$ Å[17,18,19]. In contrast to conventional BCS superconductors, it is conceivable that the amplitude and phase coherence in high-$T_c$ superconductors have different length scales since the ratio $R = 2\Delta/k_BT_c$ is no longer a constant. Recent ARPES measurements[11] suggest that $R \propto 1/x$. Thus we may expect the superconducting phase coherence length to be determined by $\eta v_F/k_BT_c$, which scales as $1/x$ on the underdoped side. An extension of our correlation length and vortex core-size measurements to underdoped samples with various doping concentrations may distinguish the two length scales since they might have different doping (x) dependences.

The observation of microscopic spatial variations in both the carrier density and the superconducting gap, and the strong correlation between them, elucidates for the first time the local inhomogeneous charge environment and its intimate relationship with superconductivity. Further exploration of this frontier may lead to a greater understanding of how high-$T_c$ superconductivity arises from doping a Mott insulator.

---

**Acknowledgements**

We acknowledge P.A. Lee and E.W. Plummer for their inspiring suggestions and encouraging comments. We also thank P.W. Anderson, A. Balatsky, D.A. Bonn, A. Castro-Neto, E. Carlson, M. Franz, L.H. Greene, X. Hu, T. Imai, B. Keimer, S.A. Kivelson, K. Kitasawa, R.B. Laughlin, D.-H. Lee, A.H. MacDonald, A. Millis, N.P. Ong, Z.-X. Shen, H.-J. Tao, X.-G. Wen, Z.-Y. Weng, N.-C. Yeh, G.-M. Zhang, and Z.-X. Zhong for helpful discussions. This work was supported by NSF, DOE, Sloan Research Fellowship, Research Corporation, the Miller Institute for Basic Research, and Grant-in-Aid for Scientific Research on Priority Area and a COE Grant from the Ministry of Education, Japan.

**Figure 1** Topographic image and associated integrated LDOS map measured on the surface (BiO plane) of an optimally oxygen doped, nominally pure single crystal of $Bi_2Sr_2CaCu_2O_{8+x}$. **a**, Constant-current-mode, topographic image (150 ×150 Å) of the surface BiO plane exposed after cleavage of the single crystal. In addition to the clear contrast due to the topological corrugations of the atomic structures and the well-known incommensurate structural modulation, an inhomogeneous background is also visible. In constant-current-mode, the tunnelling current is exponentially related to the tip-sample distance and is also proportionally related to the integrated LDOS. Therefore, the constant current topographic image provides the convolved information of both the topology and the LDOS of the crystal surface. **b**, To reveal the inhomogeneous background more clearly, we use Fourier filtering to remove the contrast due to the two well-ordered topological structures mentioned above. The variation of the integrated LDOS, which is seen as an inhomogeneous background in **a**, is now clearly displayed. A brighter colour represents a larger magnitude of the integrated LDOS.

**Figure 2** A side-by-side comparison of an integrated LDOS map and its corresponding superconducting gap map, including their associated statistical results. **a**, A 600 × 600 Å LDOS map obtained with the same technique used in Fig.1b on a single crystal $Bi_2Sr_2CaCu_2O_{8+x}$ doped with a very dilute concentration of Zn atoms. (The Zn concentration measured by STM is 0.2%[2].) The crystal has a superconducting transition temperature of 84 K, with a transition width of 4 K. The red box in the upper right corner of **a** frames a 150 ×150 Å region for easy comparison with Fig. 1b. It clearly displays an inhomogeneous structure similar to that observed in Zn-free crystals in Fig. 1b. **b**, Superconducting gap map, obtained simultaneously with the integrated LDOS map on the same location, showing the spatial variation of the superconducting energy gap. The local gap values are extracted from the corresponding local differential conductance spectra. Reverse colour coding (higher-intensity corresponding to a smaller gap magnitude) is assigned to the map for easily visualizing its correlation with the integrated LDOS map. **c** and **d** are the histograms showing the statistical distributions of the integrated LDOS and the magnitude of the superconducting gap. Each of them exhibits a Gaussian-like distribution (fitting function displayed in red). The fit of the gap distribution (42 meV mean; ~20 meV FWHM) shows it to be slightly skewed. **e** and **f** show the azimuthally averaged results of the 2D-auto-correlation analysis on the integrated LDOS map and superconducting gap map respectively. Fitting both data sets with a simple exponential decay model (red lines) results in decay lengths of ~14 Å, demonstrating the short length scale of the variations. The imperfections of the fit imply that a more complex model might be needed. **g**, The superconducting gap is spatially correlated with the LDOS as characterized by this 2D-cross-correlation function. It has a pronounced centre peak that can be fit with a Gaussian function.

**Figure 3** Spatial variation of the tunnelling differential conductance spectrum. To account for variations in the tunnelling junction, the spectra are normalized to a constant tip-sample separation. **a**. A three-dimensional rendering of the tunnelling spectra along a 200 Å line, showing detailed variations of the LDOS, the energy gap, and their correlation. Notice that the coherence peak heights vary more dramatically than the gap widths. **b**. The same data from a bird's-eye view. To demonstrate the gap variation, the black dotted lines trace the positions of the coherence peaks. Together with the gap map, one can see that the gap varies less rapidly on a "patch" than at its edge. **c**. Five characteristic spectra taken at the positions correspondingly marked in both Fig. 2a, b showing the correlation between the superconducting gap and the integrated LDOS. Curves 1 and 2 are taken at nearby positions in a very dark area on the integrated LDOS map, where the integrated LDOS is very small. The low differential conductance and the absence of a superconducting gap are indicative of insulating behaviour. Curve 3 has a large gap of 65 meV, with low coherence peaks, resembling the spectral line-shape measured in oxygen underdoped samples[4,5]. The integrated value of the LDOS at position 3 is small but larger than those in curves 1 and 2, as it is in a slightly brighter area. Curve 4 comes from a still brighter area. It has a gap value of 40 meV, which is close to the mean value of the gap distribution. The coherence peaks are sharper and higher, resembling the spectral behaviour of a sample with optimal oxygen doping[4,5]. Curve number 5, taken at the position with the highest integrated LDOS, has the smallest gap value of 25 meV with two very sharp coherence peaks, resembling the spectral behaviour of an oxygen over-doped crystal[4,5].

**Figure 4** A scatter plot of the superconducting gap versus integrated LDOS. The 16,384 data points of the corresponding integrated LDOS map and the gap map (Fig.2a,b) are plotted here in colour. A darker colour represents a larger number of data points with the same integrated LDOS and the same gap magnitude. For comparison, a set of data from ARPES measurements on various $Bi_2Sr_2CaCu_2O_{8+x}$ single crystals with different oxygen doping concentrations is superimposed onto the plot. The gap values of ARPES data are the maximum values of the $d$-wave gap at $(\pi, 0)$ of the Brillouin zone. They are presented as open circles in the plot and the corresponding doping concentration scale is placed on top of the plot. The solid black line is the linear fit to both the STM and the ARPES data.

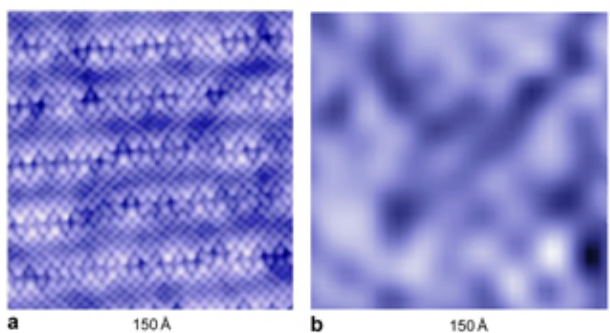

Figure 1

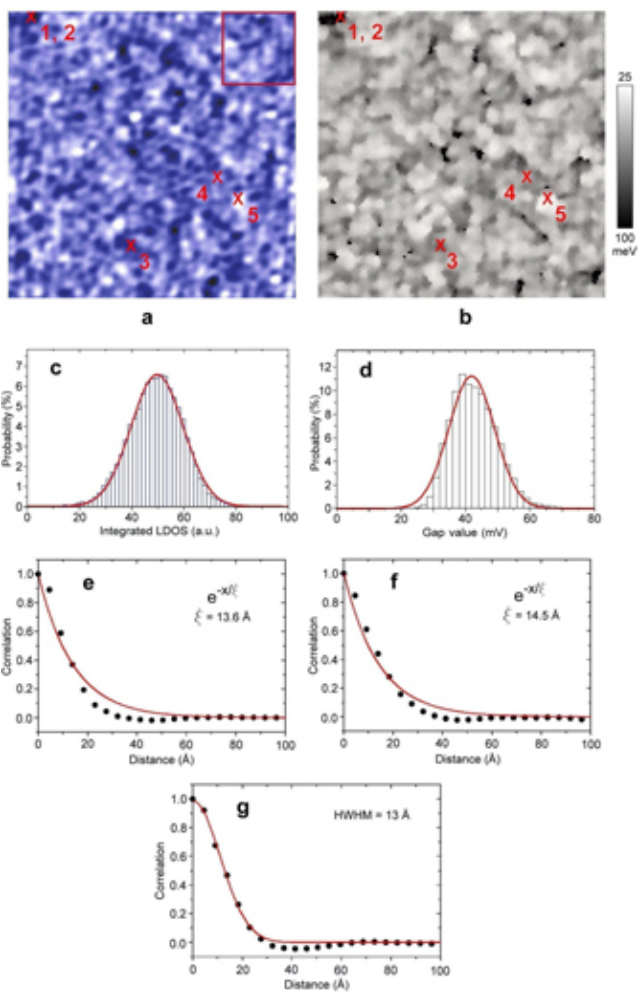

Figure 2

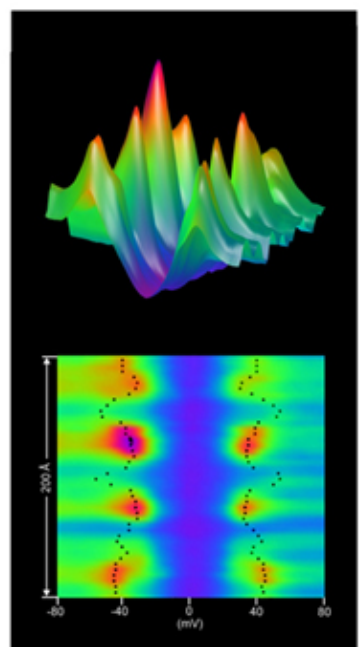

Figure 3

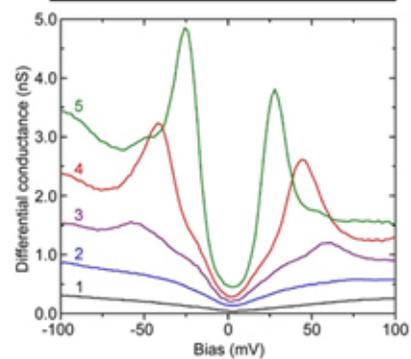

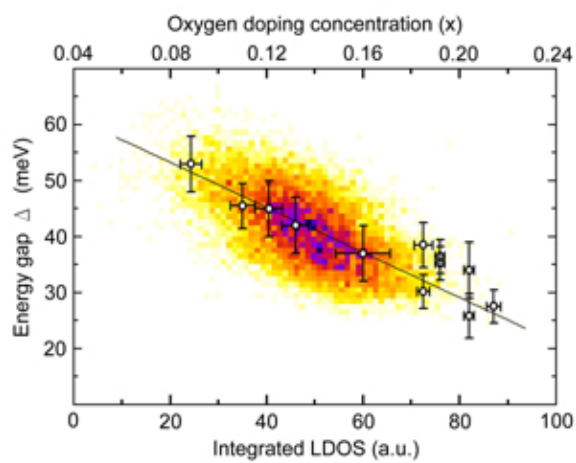

Figure 4